\documentclass[letter]{aa}

\usepackage{graphicx}
\usepackage{txfonts}
\usepackage{placeins}
\usepackage[colorlinks=true,linkcolor=blue,citecolor=blue,urlcolor=blue]{hyperref}

\begin{document}

\title{Populations of Neutron Star Ultraluminous X-ray Sources}
\subtitle{Mind your $b$'s and $B$'s}
\titlerunning{Populations of Neutron Star Ultraluminous X-ray Sources}
\authorrunning{Kovlakas et al.}

\author{
    Konstantinos Kovlakas\inst{\ref{inst:k1},\ref{inst:k2}} \and 
    Devina Misra\inst{\ref{inst:dm}} \and
    Roberta Amato\inst{\ref{inst:inaf}} \and
    Gian Luca Israel\inst{\ref{inst:inaf}}
}

\institute{
    Institut d'Estudis Espacials de Catalunya (IEEC), Edifici RDIT, Campus UPC, 08860 Castelldefels (Barcelona), Spain \label{inst:k1}
        \and 
    Institute of Space Sciences (ICE, CSIC), Campus UAB, Carrer de Magrans, 08193 Barcelona, Spain \label{inst:k2}\\\email{kovlakas@ice.csic.es}
        \and
    Department of Physics, Norwegian University of Science and Technology, NO-7491 Trondheim, Norway \label{inst:dm}
        \and
    INAF - Osservatorio Astronomico di Roma, via Frascati 33, I-00078 Monte Porzio Catone, Italy \label{inst:inaf}
}

\date{Received December 1, 2024}

\abstract{
Ultraluminous X-ray sources (ULXs) with neutron star (NS) accretors challenge traditional accretion models, and have sparked a debate regarding the role of geometrical beaming and strong magnetic fields ($B$). The reduction of the Thomson cross-section in the presence of strong $B$, leads to a modification of the Eddington limit, and therefore is expected to affect significantly the observational appearance of NS-ULXs.
We investigate the role of this modification using population synthesis models, and explore its effects on the X-ray luminosity functions, spin-up rates, and outflow energetics of the observed NS-ULXs.
Our results show that the new prescription allows NS-ULXs to achieve super-Eddington luminosities with milder beaming compared to before, improving the agreement with observations.
In addition, it broadens the range of spin-up rates allowing for more diverse conditions in NS-ULXs in terms of accretion rates and magnetic fields.
More importantly, the reduced beaming increases the likelihood of observing the NS-ULXs within wind-powered nebulae such as NGC\,5907\,ULX-1.
Our findings highlight the necessity of taking into account $B$ effects independently of the approach: geometrical beaming or strong $B$, and call for magnetospheric accretion prescriptions that can be integrated in population synthesis codes.
}


\maketitle


\section{Introduction}

Ultraluminous X-ray sources (ULXs) are accreting compact objects, distinct from active galactic nuclei, with luminosities ranging from $10^{39}$ to $10^{42}\rm\,erg\,s^{-1}$ \citep[e.g.,][]{Kaaret17,King23}. Based on the Eddington limit, assuming pure hydrogen material,
\begin{equation}
    L_{\rm Edd} = 1.26 \times 10^{38}\rm\,erg\,s^{-1} M_1,
    \label{eq:Ledd}
\end{equation}
where $M_1$ is the mass of the accretor in solar units ($M_\odot$), ULXs could represent accreting black holes (BHs) with $M_1{\sim}10{-}10^4$. Though, their association with star-forming regions suggests they are X-ray binaries \citep[e.g.,][]{Zezas02}, in which accretors have lower masses \citep[${\lesssim}20\rm\,M_\odot$;][]{CygnusX1}. Consequently, the majority of ULXs are now thought to be super-Eddington accretors.

The \citet{Shakura73} model for super-critical accretion onto BHs, where the mass-transfer rate $\dot{M}_{\rm tr}$ exceeds the Eddington rate $\dot{M}_{\rm Edd}$, describe an accretion disk that is locally Eddington-limited at all radii and becomes geometrically thick near the accretor. This model predicts strong outflows, and a bolometric luminosity higher than the $L_{\rm Edd}$,
\begin{equation}
    L_{\rm bol} = L_{\rm Edd}
    \begin{cases}
        1 + \ln \dot{m} & \dot{m} > 1 \\
        \dot{m} & \dot{m} \leq 1
    \end{cases},
    \label{eq:lbol}
\end{equation}
where $\dot{m}{=}\dot{M}_{\rm tr}/\dot{M}_{\rm Edd}$ is the Eddington ratio, and $L_{\rm Edd}{=}0.1 \dot{M}_{\rm Edd} c^2$ adopting accretion efficiency $0.1$ (\citealt{Frank02}. For completeness, the sub-Eddington regime ($\dot{m}{<}1$) is also included. We note that Eq.~\eqref{eq:lbol} does not take into account the role of advection \citep[e.g.,][]{Beloborodov98,Chashkina19} which may decrease the bolometric luminosity by a factor of ${\sim}30{-}40\%$ \citep[e.g,][]{Lipunova99,Poutanen07}.
Still, luminosities of ${\gtrsim}10^{40}\rm\,erg\,s^{-1}$ cannot be accounted for by the $\ln\dot{m}$ term. Instead, they are better explained by the collimation of the radiation escaping through a funnel formed by the outflows \citep{King01}. As a result, the observed luminosity under the isotropic emission, $L_{\rm iso}$, is
\begin{equation}
    L_{\rm iso} = b^{-1} L_{\rm bol},
    \label{eq:Lbol}
\end{equation}
where the beaming factor, $b$, represents the fraction of the full solid angle through which radiation escapes. The $b$ has been shown to correlate with the $\dot{m}$ \citep{King09},
\begin{equation}
    b = \begin{cases}
            1 & \dot{m} \leq 8.5, \\
            \left(8.5/\dot{m}\right)^2 & \dot{m} > 8.5
        \end{cases},
    \label{eq:bfactor}
\end{equation}
allowing for extreme luminosities without invoking extreme BH masses (e.g., $10^{42}\rm\,erg\,s^{-1}$ for $M_1{=}30$ and $\dot{m}{=}60$).

The discovery of pulsating ULXs \citep[PULXs;][]{Bachetti14} confirmed the existence of neutron star (NS) ULXs, often exhibiting highly super-Eddington luminosities. The most extreme example, NGC\,5907\,ULX-1, reached $L_{\rm iso}/L_{\rm Edd}{\sim}1000$ \citep{Israel17}, which under the beaming scenario, corresponds to small $b{<}10^{-2}$. Such low $b$ values are inconsistent with the sinusoidal profiles in PULXs \citep[e.g.,][]{Kaaret17}, and are not reproduced by numerical simulations \citep[e.g.,][]{Abarca21}. This has prompted investigations into alternative scenarios, primarily focusing on the role of the ``usual suspect'' in NSs, the magnetic field \citep[e.g.,][]{Eksi15}.

In the presence of strong magnetic fields, the Thomson cross-section is reduced for photon energies below the cyclotron energy \citep{Herold79}, increasing the Eddington limit to a critical luminosity \citep[e.g.,][]{Paczynski92,DallOsso15,Brightman18},
\begin{align}
    L_{\rm crit} &\approx 2 B_{\rm 12}^{4/3} L_{\rm Edd},
    \label{eq:Lcrit}
\end{align}
where $B_{12}$ is the dipolar magnetic field in units of $10^{12}\rm\,G$. Consequently, NS-ULXs could be sub-Eddington accretors with $B_{12}{\sim}10{-}1000$ without beamed emission. However, this interpretation relies on highly magnetised NSs, which could push many PULXs in the propeller regime, where infalling matter is expelled rather than accreted.

Whether the observational appearance of PULXs can be explained by  beaming or extreme $B$'s is a matter of debate since their discovery. Arguments in favour of the beaming are:
(i) 
    The high-$B$ scenario requires $B_{12}{\gtrsim}100$ \citep[e.g.,][]{Lasota23}, higher than the typical values in X-ray binaries and constraints in PULXs \citep[e.g.,][]{Walton18,Middleton19};
(ii) 
    Collimated emission in NS-ULX simulations \citep[e.g.,][]{Mushtukov19,Inoue23}, and its consequences for the spectra of ULXs \cite[e.g.,][]{Poutanen07,King09};
(iii) 
    The extension of the X-ray luminosity function (XLF) to the ULX regime, reproduced by population synthesis studies employing beaming, in star-forming galaxies \citep[e.g.,][]{Misra23}, including BH-dominated low-metallicity galaxies \citep[e.g.,][]{Wiktorowicz19}, and in NS-dominated passive galaxies agreeing with ULX demographic studies \citep[e.g.,][]{Kovlakas20}.

Conversely, the high-$B$ scenario ($B_{12}{\gtrsim}100$) has been supported by the following points:
(i)
    Expanding nebul\ae{} near ULXs have been found to require mechanical power, in terms of outflow kinetic power ($L_{\rm kin}$), similar to the observed luminosity of the ULXs \citep[e.g., Holmberg\,II\,X-1;][]{Pakull2002}, suggesting that the inferred luminosities are not overestimated due to strong beaming (e.g., $L_{\rm kin}{\sim}1.3{\times}10^{41}\rm\,erg\,s^{-1}$ in NGC\,5907\,ULX-1; \citealt{Belfiore2020});
(ii) 
    The orbital decay observed in M82\,X-2 points at $\dot{M}_{\rm tr}{\gg}\dot{M}_{\rm Edd}$, while the mass available to the accretor is sufficient to explain its luminosity \citep[e.g.,][however see \citealt{King21}]{Bachetti2022}, as well as mild beaming hinted by the long-term spin-down rate of the system \citep{Liu24};
(iii)
    Extreme beaming implies a large population of hidden ULXs, challenging to reproduce in population synthesis models which often force a $b$ limit \citep[e.g., ${>}0.0032$;][]{Wiktorowicz19,Misra23};
(iv) Strong beaming precludes the detection of pulsations due to the reflections of X-ray photons in the geometrically-thick inner accretion wall \citep[e.g.,][]{Mushtukov21,Mushtukov23}.

It has been suggested that the observational appearance of PULXs may be explained by the combined effect of $B$ on $L_{\rm Edd}$, the presence of multipolar components, and moderate beaming \citep[e.g.,][]{Israel17,Erkut20}. However, to date, there is no comprehensive prescription for the magnetic field configuration of NSs, and magnetospheric accretion that is applicable to both sub-Eddington and super-Eddington regimes. Consequently, the models of \citet{Shakura73,Poutanen07,King09} are commonly employed in observational \citep[e.g.,][]{Middleton19} and theoretical studies \citep[e.g.,][]{Lasota23}, as well as in population synthesis models \citep[e.g.,][]{Wiktorowicz19,Misra23,Misra24}.

In this Letter, we evaluate whether the combination of the modified $L_{\rm Edd,B}$ and beaming reconciles the observations discussed above, using population synthesis models. Finally, we highlight the importance of integrating in the models up-to-date magnetospheric accretion prescriptions towards a self-consistent framework for the study of NS-ULXs.


\section{Methodology}

\begin{figure*}
    \centering
    \includegraphics[width=0.98\linewidth]{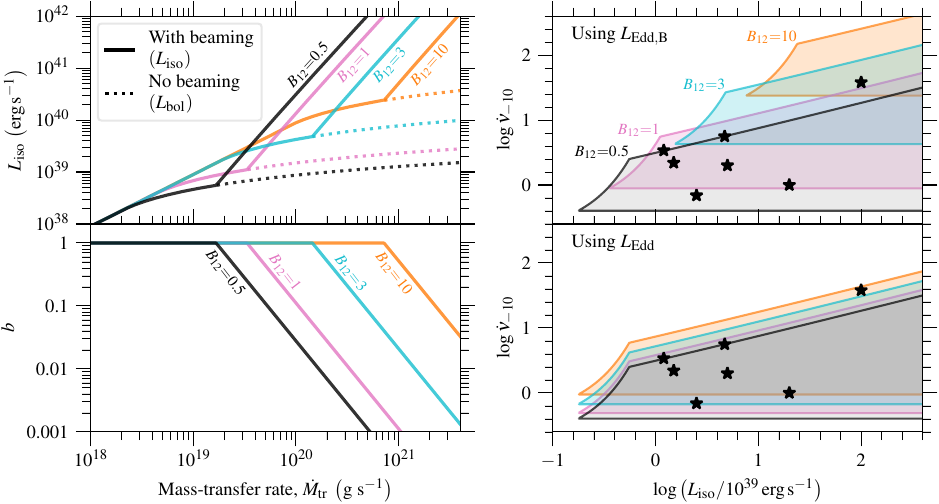}
    \caption{Effects of beaming and $L_{\rm Edd,B}$ for a $1.4\rm\,M_\odot$ NS of varying $B$ (see coloured text) on the apparent luminosity $L_{\rm iso}$ (\textit{upper left}) with/without beaming (solid and dotted lines, respectively), and $b$ (\textit{bottom left}) as a function of the $\dot{M}_{\rm tr}$. We also compare the $\dot{\nu}$ ranges 
    (accretion rate between $\dot{M}_{\rm Edd}$ and $\dot{M}_{\rm tr}$)
    as a function of the $L_{\rm iso}$ using the modified (\textit{upper right}) and classical (\textit{lower right}; increasing $B$ from bottom to top) $L_{\rm Edd}$ against the values of observed PULXs (asterisks; \citealt{King19}).}
    \label{fig:effects}
\end{figure*}

In the following paragraphs we describe the calculation of the observable quantities ($L_{\rm iso}$, $L_{\rm kin}, \dot{\nu}$) in simulated systems, as well as the setup of our population synthesis models.

\subsection{Observables}

The $B$-dependent critical luminosity acts as a new Eddington limit, only if $L_{\rm crit}{>}L_{\rm Edd}$ \citep{Paczynski92}:
\begin{equation}
    L_{\rm Edd,B} =
    \max\left\{1, 2 B_{12}^{4/3}\right\} L_{\rm Edd},
    \label{eq:newEdd}
\end{equation}
combining Eqs. \eqref{eq:Ledd} and \eqref{eq:Lcrit}. Then,
Eqs. \eqref{eq:Ledd}--\eqref{eq:bfactor} are applied to estimate the apparent luminosity, $L_{\rm iso}$, for both BHs and NSs, but now the $\dot{m}$ is determined relative to the $L_{\rm Edd,B}$ for NSs.

The mechanical feedback ($L_{\rm kin}$) is challenging to model due to its dependence on the outflow rate and speed, both of which vary as functions of the launching radius. This distribution is shaped by the accretion geometry and $\dot{M}_{\rm tr}$.
To estimate $L_{\rm kin}$ at an order-of-magnitude level, we assume that matter is accreted at approximately the Eddington rate \citep[however it can exceed it by a factor of 6 in NSs;][]{Basko76,Kaaret17}, while the rest of the transferred mass is expelled with speed ${\approx}0.2\,c$ \citep[e.g.,][]{Pinto16}:
\begin{equation}
    L_{\rm kin} \approx \frac{1}{2} \left(\dot{M}_{\rm tr} - \dot{M}_{\rm Edd}\right) \left(0.2\,c\right)^2 = \frac{\dot{m}-1}{5} L_{\rm Edd}.
\end{equation}

The spin-up rate $\dot{\nu}$ in NSs is the result of the torque from the accreted material at rate $\dot{M}$ (at $R_{\rm M}$):
\begin{equation}
    \dot{\nu} = \frac{\dot{M}\left(G M R_{\rm M}\right)^{1/2}}{2\pi I},
\end{equation}
where $I{\approx}10^{45}\rm\,g\,cm^{2}$ is the moment of inertia. The $R_{\rm M}$ is the magnetospheric radius,
\begin{equation}
    R_{\rm M} = \left(\frac{\mu^4}{2GM\dot{M}^2}\right)^{1/7},
\end{equation}
where $\mu{=}BR^3$ is the magnetic moment with dipolar field $B$ and radius $R$.
From these we get
\begin{align}
    R_{\rm M} &= 3.24{\times}10^{8}\;
        \dot{M}_{17}^{-2/7} M_{1}^{-1/7} B_{12}^{4/7} R_6^{12/7}\,\mathrm{cm}, \label{eq:Rm_simple}\\
    \dot{\nu} &= 3.30{\times}10^{-12}\; 
        \dot{M}_{17}^{6/7} M_{1}^{-3/7} B_{12}^{2/7} R_6^{6/7}\,\mathrm{s^{-2}}, \label{eq:nudot}
\end{align}
where the accretion rate $\dot{M}_{17}$ (at $R_{\rm M}$), magnetic field $B$, radius $R_6$, and mass $M_{1}$, are in units of $10^{17}\,\rm g\,s^{-1}$, $10^{12}\,\rm G$, $10^6\rm\,cm$, and $1\,M_\odot$, respectively. We notice that Eq.~\eqref{eq:Rm_simple} is consistent with \citet{Shapiro83}, but with larger scaling than the expressions in \citet{King17} and \citet{Lasota23}, as well as a typo in the exponent of the mass term in the latter.

\subsection{Population synthesis models}

The $L_{\rm Edd}$ corresponding to characteristic masses of compact objects in X-ray binaries may manifest as breaks in XLFs \citep[e.g.,][]{Kaaret17}. Moreover, beaming ``carves'' the XLFs near these limits and extends them, by shifting the portion of the population that is beamed towards the line of sight to higher luminosities, while concealing the rest. As a result, beaming prescriptions can be tested at the population level by comparing population synthesis models with observed XLFs in star-forming galaxies (all PULXs have been found in spiral galaxies; \citealt{King23}).

Different codes and choices of stellar evolution parameters can influence the NS/BH ratio, mass-transfer rates, etc., leading to varying XLF predictions. Here our goal is not to constrain these parameters, but rather to explore the effects of beaming and $L_{\rm Edd,B}$ while ensuring that the results are not an artifact of the specific modelling choices.
For this reason, we employ two drastically different approaches: (i) a parametric binary population synthesis code, \texttt{COSMIC} v.3.4.17 \citep{Breivik20}, using the default stellar evolution parameters, and (ii) \texttt{POSYDON} v.1 \citep{Fragos23}, which incorporates up-to-date physics and detailed stellar tracks with \texttt{MESA} \citep{Paxton11}, using parameters reproducing the XLF \citep[c.f., model 44 in][]{Misra23,Misra24}. We evolve $10^7$ massive binaries in a constant $100\rm\,Myr$ star-formation scenario, and select systems with $\dot{M}_{\rm tr}$ larger than 1\% of the classical $L_{\rm Edd}$ to ensure that the full ULX population is included in the samples (including sub-Eddington ULXs with massive BHs).
Focusing on the shape, rather than the normalisation, of the XLFs, we scale the populations to match the measured scaling of ULXs with star-formation rate \citep[${\sim}0.5$ ULXs per $1\rm\,M_\odot\,yr^{-1}$;][]{Kovlakas20,Lehmer21}. Since \texttt{POSYDON} does not evolve the magnetic field of NSs, we sample $B$ values from the resulting distribution in the \texttt{COSMIC} population: $\log\left(B / {\rm1\, G}\right) \sim \mathcal{N}(12.22, 0.64)$ with maximum at $13.79$. This distribution is consistent with the expectation that PULXs are high-mass X-ray binaries with age shorter than the $B$-decay timescale \citep[e.g.,][]{Revnivtsev16}.


\section{Results}

\begin{figure*}
    \centering
    \includegraphics[width=0.49\linewidth]{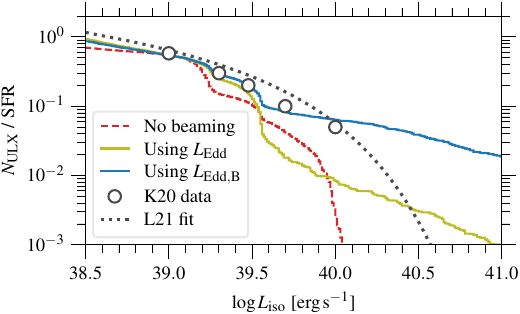}
    ~
    \includegraphics[width=0.49\linewidth]{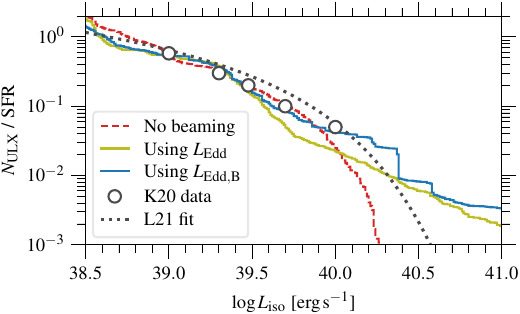}
    \caption{Synthetic X-ray luminosity functions using \texttt{COSMIC} (\textit{left}) and \texttt{POSYDON} (\textit{right}), in the case of no beaming (red dashed), and beaming with the traditional (green), or modified Eddington limit (blue). The populations include both BH and NS-ULXs, and are compared against observational constraints from the XLF fit of \citet[][L21 in legend]{Lehmer21} for nearby solar-metallicity star-forming galaxies (black), and ULX demographic data (open circles) from \citet[][K20 in legend]{Kovlakas20} using larger galaxy samples \citep{Kovlakas21}.}
    \label{fig:xlf}
\end{figure*}

In the left panels of Fig.~\ref{fig:effects} we show the $L_{\rm iso}$ (upper panel) and $b$ (lower panel) as a function of the $\dot{M}_{\rm tr}$, for $M_1{=}1.4$ and various values of $B_{\rm 12}$. We omit showing the effect of NS mass as it is negligible given its small range (${\sim}1{-}2\rm\,M_\odot$). For $B_{12}{=}0.5$ the results are numerically equivalent to using the classical $L_{\rm Edd}$, as the $B$-dependence activates for $B_{12}{>}0.6$ (Eq.~\eqref{eq:Lcrit}). We find that NS-ULXs with moderately high $B_{12}{=}10$ can reach luminosities up to $10^{41}\rm\,erg\,s^{-1}$ with mild beaming (${>}0.1$). Without beaming (dashed lines), significantly higher $B_{12}$ is necessary to account for NS-ULXs with $L_{\rm iso}{\approx}10^{41}\rm\,erg\,s^{-1}$.

In Eq.~\eqref{eq:nudot}, the $\dot{M}$ is the accretion rate at the $R_{\rm M}$, making $R_{\rm M}$ and $\dot{M}$ co-dependent. Therefore, they cannot be calculated without exact knowledge of the accretion geometry, and are only constrained via measurements of $\dot{\nu}$ and $B$. However, we can estimate a conservative range for the $\dot{\nu}$ considering $\dot{M}$ values between $\dot{M}_{\rm Edd}$ (the minimum rate for a super-Eddington NS) and $\dot{M}_{\rm tr}$ (the case where all transferred mass is accreted). The right panels of Fig.~\ref{fig:effects} depict these ranges for $M_1{=}1.4$ and various values of $B_{12}$, considering the $B$-dependent (upper) and the classical $L_{\rm Edd}$ (lower). In both cases, the regions do not include spin-up rates corresponding to the propeller regime, that is $R_{\rm M}$ is less than the corotation radius adopting the fastest spin observed in PULXs \citep[$0.42\rm\,s$;][]{Furst16} to remain conservative. The points are the measured values from PULXs taken from \citet{King19}, and lie within the regions defined by the model using both prescriptions of the $L_{\rm Edd}$.

In Fig.~\ref{fig:xlf} we present the synthetic XLFs, incorporating both BH and NS-ULXs. Employing the $B$-dependent $L_{\rm Edd,B}$ (blue lines), instead of the classical $L_{\rm Edd}$ (green lines), enhances the contribution of NS-ULXs. This is attributed to the reduced $b$, which reveals a larger fraction of the systems. The increased presence of NS-ULXs improves the agreement with the observations, despite the difference between the \texttt{COSMIC} and \texttt{POSYDON}, particularly in the slope at the less-constrained ${>}10^{40}\rm\,erg\,s^{-1}$ regime, in which additional processes might be at play \citep[e.g., neutrino energy loss][]{Mushtukov18}.

\begin{figure*}
    \centering
    \includegraphics[width=0.98\linewidth]{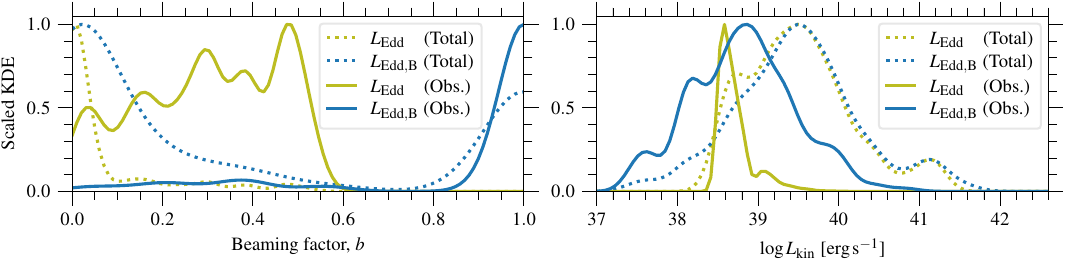}
    \caption{Kernel density estimates (scaled to unity for visibility) of the beaming factor (\textit{left}) and mechanical feedback (\textit{right}) in the total (dashed) and observed (solid) population of NS-ULXs from the \texttt{POSYDON} model, using the classical (green) or modified Eddington limit (blue).
    }
    \label{fig:beamkin}
\end{figure*}

In the left panel of Fig.~\ref{fig:beamkin} we compare the beaming factors of NS-ULXs ($L_{\rm iso}{\in}[10^{39},10^{42}\rm\,erg\,s^{-1}$) for the classical (green) and $B$-dependent $L_{\rm Edd}$ (blue), both of the total (dashed curves) and the observed (weighted by $b$; solid curves) synthetic populations.
The classical $L_{\rm Edd}$ implies a highly-beamed total population, with a moderately-beamed ($b{\sim}0.1{-}0.5$) observed subpopulation with luminosities in the $10^{39}{-}10^{40}\rm\,erg\,s^{-1}$ range (cf. Fig.~\ref{fig:effects}). In contrast, the use of $L_{\rm Edd,B}$ results in more diverse populations. About half of NS-ULXs are moderately beamed ($b{\sim}0.05-0.5$), with the rest representing the majority of the observed population with mild beaming ($b{>}0.7$) and luminosities spanning the full range of PULXs for $B_{12}{\sim}0.5{-}10$ (cf. Fig.~\ref{fig:effects}).

The order-of-magnitude estimate for the kinetic power of outflows in NS-ULXs can reach up to $10^{42}\rm\,erg\,s^{-1}$. Depending mainly on the $\dot{M}_{\rm tr}$, it is not affected by the $L_{\rm Edd}$ prescription for the total population (dashed lines in Fig.~\ref{fig:beamkin}; right panel). However, when we account for the observed fraction of NS-ULXs (not beamed out of the line of sight), the distribution of $L_{\rm kin}$ shifts downward by one and two orders of magnitude using the $L_{\rm Edd,B}$ and $L_{\rm Edd}$, respectively.

\section{Discussion}

The increased Eddington limit, $L_{\rm Edd,B}$ allows for both sub- and super-Eddington NS-ULXs without beaming, as well as super-Eddington NS-ULXs with milder beaming as $B$ increases. This contrasts with the the classical $L_{\rm Edd}$ leading to all NS-ULXs being strongly beamed.
Interestingly, sub-Eddington NS accretors do not need to exhibit $B_{12}{\gtrsim}100$ to explain the PULX luminosities, since for $L_{\rm Edd,B}{=}10^{39}{-}10^{41}\rm\,erg\,s^{-1}$ we have $B_{12}{\sim}2{-}70$.

The effect of the magnetic field in the critical luminosity of NS-ULXs enhances their contribution in the XLFs, and helps smooth out prominent breaks \citep[e.g.,][]{Kaaret17}, matching better with observed XLFs \citep[e.g.,][]{Lehmer21} and ULX demographics \citep[e.g.,][]{Kovlakas20}. This is achieved with magnetic fields in the observed range of X-ray binaries ($B_{12}{\sim}0.1{-}10$).
However, since our modelling focuses on the general properties of NS-ULXs and the effects on the XLFs, we do not exclude the possibility of multi-polar component magnetic fields that might be required to explain all observables in individual PULXs \citep[e.g.,][]{Israel17}.

The inferred mechanical power of $L_{\rm kin}{\approx}1.3{\times}10^{41}\rm\,erg\,s^{-1}$ for the nebula near NGC\,5907\,ULX-1 \citep{Belfiore2020} is comparable to the apparent ULX luminosity. This often serves as an argument against beaming, however, it remains uncertain whether the mechanical power must be similar to the bolometric luminosity. Our order-of-magnitude estimates of $L_{\rm kin}$ can reach even higher values provided high $\dot{M}_{\rm tr}$ values, independently of beaming. More importantly, using the classical $L_{\rm Edd}$ leads to small $b$, which makes it unlikely to find NS-ULXs close to wind-powered nebul\ae{}. This is evident from the the shift of two orders of magnitude in the total and observed distribution in Fig.~\ref{fig:beamkin}. Conversely, the lower $b$ in the modified $L_{\rm Edd,B}$ results in a shift of only one order of magnitude, increasing the chances of observing systems like NGC\,5907\,ULX-1. Systematic searches of wind-powered nebul\ae{}, with or without detection of the PULXs powering them, can measure this shift, and put constraints in beaming and outflow models of PULXs.

Using the classical $L_{\rm Edd}$ results in a range of $\dot{\nu}$ values that just covers the observed values (Fig.~\ref{fig:effects}; lower right), with half of them indicating spin-up rates close to the maximum possible for $B_{12}{\lesssim}10$. This requires either extreme $B$ values, or the majority of the transferred mass to reach the magnetospheric radius ($\dot{M}{\approx}\dot{M}_{\rm tr}$), potentially leading to reduced outflows. In this analysis, we neglect disk-star coupling that might introduce additional angular momentum losses, thereby further reducing the spin-up rate. On the other hand, the $B$-dependent $L_{\rm Edd,B}$ reduces the $\dot{m}$, shifting both the apparent luminosities, and the locus of super-Eddington sources in the diagram, resulting in a broader range of spin-up rates. Under this scenario, NS-ULXs may exhibit a wider variety of spin-up behaviours, with moderate $B$.
Discovery of PULXs with higher spin-up rates could rule out the classical $L_{\rm Edd}$ scenario.

We highlight the importance of incorporating a more detailed treatment of the effects of magnetic fields in NS-ULXs, especially with respect to the $L_{\rm Edd}$. 
Towards a self-consistent framework of NS-ULXs, we need parameter studies (investigating kick velocity distributions, mass-transfer prescriptions; e.g., \citealt{ElMellah2019}, etc.) with population synthesis codes integrating results from magnetospheric accretion simulations. The latter can provide more realistic prescriptions for the bolometric luminosity and $b$ in NS-ULXs \citep[e.g.,][]{Vasilopoulos21}, which are often extrapolated \citep[e.g.,][]{Wiktorowicz19,Misra24} from the BH-ULX population \citep{King09}, although they are expected to be qualitatively similar \citep[e.g.,][]{King23}. Furthermore, the dependence of $R_{\rm M}$ and $\dot{M}(R_\mathrm{M})$ \citep[e.g.,][]{Chashkina17} and of the critical luminosity (see \citealt{Mushtukov15} and \citealt{Brice21} for dipolar and multipolar fields, respectively) on the accretion geometry points at the necessity of prescriptions applicable to population synthesis models.
Despite, important efforts towards understanding accretion in NS-ULXs \citep[e.g.,][]{Kuranov20,Mushtukov24}, the effect of $B$ in $L_{\rm Edd}$ in simulations of highly-magnetised NSs remains to be explored \citep[see discussion in][]{Inoue24}.
In the present work, we show that this effect leads to markedly different results at the population level, even in the case the BH-ULX prescriptions are broadly valid.

The above mentioned improvements will allow future population synthesis codes to self-consistently model mass transfer and angular momentum loss \citep[e.g.,][]{Misra20}, NS spin-up and disk-star coupling \citep[e.g.,][]{Kluzniak07}, orbital evolution \citep[e.g.,][]{Chen24}, as well as emission and detectability of pulsations \citep[e.g.,][]{King20}. This is necessary to test models using all available data from individual ULXs and PULXs, and at a population level.

Finally, using Eqs. \eqref{eq:Ledd}-\eqref{eq:newEdd}, we provide practical limits and formul\ae{} for observed sources under the case of canonical NS mass ($1.4\,M_\odot$) and radius ($10^6\rm\,cm$).
If there is evidence of beaming, then for $B_{12}{<}0.6$, ULX luminosities require $b<0.6$. This limit increases with $B$ until $B_{12}{\approx}0.9$ where $b$ approaches unity. In the weak $B$ case, the beaming factor can be constrained by the apparent luminosity (solving eq.~6 in \citealt{Kovlakas22} for $b$):
\begin{equation}
    b_{\rm min} \approx \left(5.5{\times}10^{38}\,\mathrm{erg\,s^{-1}} / L_{\rm iso}\right)^{8/9}.
\end{equation}
Evidence of beaming exceeding this value ($b_{\min}{<}b{<}1$) can be used to estimate the $B$:
\begin{equation}
    B_{12} =\left[\dfrac{ b L_{\rm iso} } {3.5{\times}10^{38}\,{\rm erg\,s^{-1}}\left(3.14+\ln b^{-1/2}\right)} \right]^{3/4},
\end{equation}
whereas, lack of beaming puts a lower limit on the $B$,
\begin{equation}
B_{12} > \left(L_{\rm iso} / 1.1{\times}10^{39}\rm\,erg\,s^{-1}\right)^{3/4}.
\end{equation}


\begin{acknowledgements}
We thank the anonymous referee for their valuable comments and suggestions, which have significantly improved the quality of this manuscript.
KK is supported by a fellowship program at the Institute of Space Sciences (ICE-CSIC) funded by the program Unidad de Excelencia Mar\'ia de Maeztu CEX2020-001058-M.
DM acknowledges that this project has received funding from the European Research Council (ERC) under the European Union’s Horizon 2020 research and innovation programme (grant agreement No. 101002352, PI: M. Linares).
RA and GLI acknowledge financial support from INAF through grant ``INAF-Astronomy Fellowships in Italy 2022 - (GOG)''. GLI also acknowledges support from PRIN MUR SEAWIND (2022Y2T94C), which was funded by NextGenerationEU and INAF Grant BLOSSOM.
\end{acknowledgements}


\bibliographystyle{aa}
\bibliography{main}

\end{document}